# Ultrafast nonlocal control of spontaneous emission

Chao-Yuan Jin[1,*], Robert Johne[1,2], Milo Y. Swinkels[1], Thang B. Hoang[1,†], Leonardo Midolo[1], Peter J. van Veldhoven[1], and Andrea Fiore[1]

[1]COBRA Research Institute, Eindhoven University of Technology, P.O. Box 513, NL-5600 MB Eindhoven, the Netherlands.

[2]Max Planck Institute for the Physics of Complex Systems, Nöthnitzer Str. 38, 01187 Dresden, Germany.

**Solid-state cavity quantum electrodynamics systems will form scalable nodes of future quantum networks, allowing the storage, processing and retrieval of quantum bits, where a real-time control of the radiative interaction in the cavity is required to achieve high efficiency. We demonstrate here the dynamic molding of the vacuum field in a coupled-cavity system to achieve the ultrafast nonlocal modulation of spontaneous emission of quantum dots in photonic crystal cavities, on a timescale of ~200 ps, much faster than their natural radiative lifetimes. This opens the way to the ultrafast control of semiconductor-based cavity quantum electrodynamics systems for application in quantum interfaces and to a new class of ultrafast lasers based on nano-photonic cavities.**

---

[*] Correspondence to: c.jin@tue.nl
[†] Currently with Department of Physics, Duke University, NC, USA.

Semiconductor-based cavity quantum electrodynamics (CQED) represents a scalable platform for quantum information processing, where the radiative interaction between emitter and photon plays a key role in the generation of non-classical light states and of entanglement[1]. When a two-level emitter in the excited state interacts with the vacuum field of a cavity mode, the evolution of the system is determined by the interplay of coupling rate $g$, interaction time $T$ and cavity loss rate $\kappa$. The control of one or more of these parameters in real time allows tailoring the interaction to the desired application[2], resulting for example in entangled emitter-photon states[3], or in single-photon states with an optimized waveform for quantum networking applications[4]. In atomic CQED, this control is realized at microwave frequencies by varying the interaction time $T$ in the tens of μs range[3], and at optical frequencies by adiabatic passage techniques[4-7], which effectively allow the shaping of the coupling rate $g$ on 100-ns timescales. In superconducting circuit quantum electrodynamics, the cavity loss rate has been changed in the 100-ns timescale by the electrical control of circuit elements[8]. In semiconductor systems, based for example on quantum dots (QDs) in photonic crystal (PhC) cavities, however, these approaches are difficult to implement, due to the different energy level structure and to the faster cavity loss and emitter's decoherence rate. The ideal control method for semiconductor CQED would allow the ultrafast manipulation of the coupling rate and/or of the cavity loss rate, without directly affecting the coherent evolution of the emitter. While a variety of methods for the control of semiconductor CQED have been demonstrated, e.g. by tuning the emitter or cavity frequency using electric field[9], strain[10], or nanomechanical deformation[11,12], none of them has been shown to provide the control of radiative processes on the sub-ns timescales needed for waveform shaping or for the control of Rabi oscillations. Here we propose an approach which enables the *nonlocal* and *ultrafast* control of the coupling rate and/or the cavity loss rate in a solid-state system. We show its implementation in semiconductor QDs weakly coupled to a PhC coupled-cavity system, demonstrating the ultrafast control of the SE dynamics at optical frequencies for the first time, with a ~200 ps temporal resolution which can be further reduced to the few ps range.

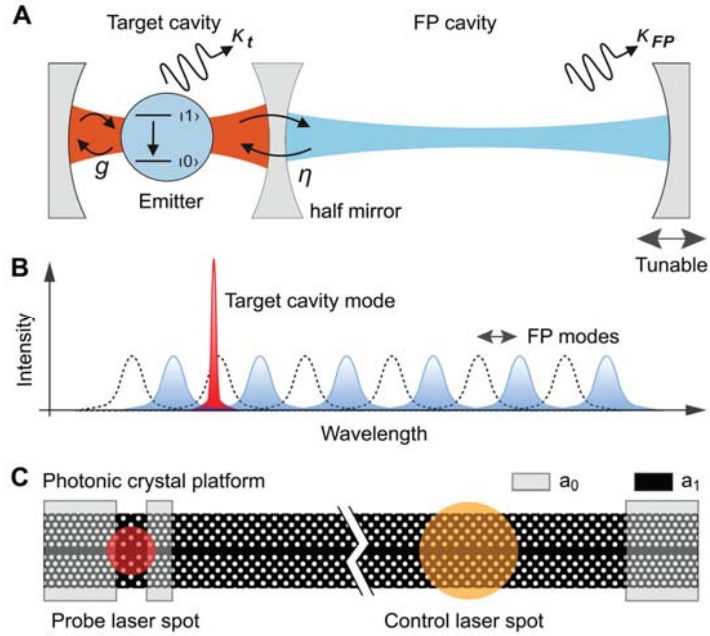

**Fig. 1.** Scheme of the nonlocal control of the emitter-photon interaction. (**A**) Two cavities with different Q-factors and mode volumes are coupled though a semi-transparent mirror. (**B**) Bringing one of the FP cavity modes into the resonance of the target cavity mode produces a redistribution of the mode field and a change of Q-factor. (**C**) A schematic image of coupled PhC cavities. The probe laser beam is located at the target cavity to generate the µPL signal while the control laser beam is focused on the FP cavity, at 30 µm distance from the target cavity.

Our approach is based on the ultrafast molding of the vacuum field, and thereby the coupling rate and cavity loss, seen by a dipole emitter sitting in a "target" cavity, by changing the resonant wavelength of an adjacent Fabry-Perot (FP) cavity coupled to the target cavity through a semi-transparent mirror (Fig. 1A). When two cavities are out of resonance, the modes of the system are well approximated by the modes of uncoupled cavities, where by design the periodic FP modes have lower quality factor (Q-factor) and larger mode volume compared to the mode of the target cavity. A change in the refractive index of the FP cavity produces a spectral shift of FP modes and brings one of them into resonance with the target cavity mode (Fig. 1B). This causes a redistribution of the vacuum field seen by the emitter, with the corresponding increase of effective mode volume and reduction of the Q-factor from the target to the FP cavity, thereby changing the emitter-field coupling rate and the loss rate. If the change of the FP spectrum is produced by an ultrafast laser pulse, by the photoexcitation of free carriers[13,14], the vacuum field responds within a timescale set by the target-FP coupling rate (typically ps), and the emitter experiences a dynamic modulation of the local density of states (LDOS) during its interaction with the cavity. A simple analysis based on coupled mode theory shows that the emitter interacts with the coupled mode with a rate $g_1 = \alpha g$, where $\alpha$ is the target-cavity component of the electric field of the coupled mode, and $g$ is the interaction rate between the emitter and the uncoupled target cavity. In the limit of weak emitter-cavity coupling of interest here, the SE rate $\gamma_1$ into one of the coupled modes, normalized by the one in the uncoupled target cavity $\gamma_t$, is given by $\frac{\gamma_1}{\gamma_t} \cong \left|\frac{g_1}{g}\right|^2 \frac{\kappa_t}{\kappa_1} = |\alpha|^2 \frac{Q_1}{Q_t}$, where $Q_1(\kappa_1)$ and $Q_t(\kappa_t)$ are the Q-factors (loss rates) of the coupled mode and uncoupled target cavity, respectively. The SE rate is affected both by the change in the coupling rate $g$ (term $|\alpha|^2$) and loss rate κ, which are controlled by the target-FP detuning. We note that also a pure $g$-modulation is possible by choosing the Q-factor of the FP cavity equal to the one of the target cavity.

We have implemented this concept using PhC cavities and semiconductor QDs as emitters. Coupled PhC cavities were previously investigated as examples of photonic molecules[15-17], for Q-factor tuning[18] and coupled-cavity quantum electrodynamics[19]. Two double-heterostructure cavities[20] are defined by slightly modifying the lattice constant along a W1 PhC waveguide from $a_0$ to $a_1 = 1.03 \times a_0$. In the first series of experiments, the change in SE rate of QDs in the target cavity was characterized in static conditions by thermo-optic tuning of the FP cavity. The FP modes shift to longer wavelength due to heating when the excitation power increases[21], which changes the detuning between two cavities (Fig. 2A). When the central FP mode crosses the target cavity mode, a decrease in emission intensity is observed and the Q-factor decreases by a maximum factor of 2.0 (Fig. 2B-C). The cavity wavelength, the Q-factor, and the micro-photoluminescence (µPL) decay time dependent on the detuning can be well fitted by the coupled mode theory (Fig. 2B-D). Taking into account the measured SE rate into the leaky modes of the PhC, a change in the SE rate in the mode by a factor of 2.7 is achieved. We note that this is larger than the relative change in Q, which is due to the redistribution of the vacuum field (term $|\alpha|^2$).

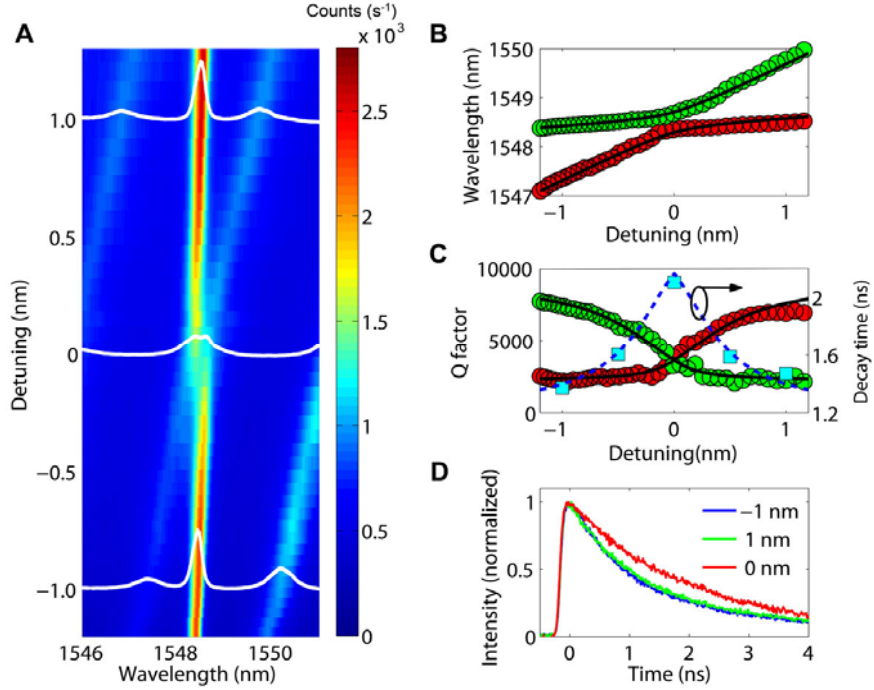

**Fig. 2.** Static modulation of the Q-factor and SE rate. (**A**) A µPL map exhibits the target cavity mode and FP modes at different detunings, corresponding to different CW laser powers on the FP cavity. Three white curves are the µPL spectra at detunings of -1.0, 0, and 1.0 nm respectively. (**B**) The wavelengths of the coupled modes as functions of the detuning. The black curves are fits using the coupled-mode theory. The observed anticrossing shows that the two cavities are at the edge between the strong and weak coupling regimes. (**C**) The Q-factors of coupled modes and the decay times of SE as functions of the detuning. The black and blue dashed curves are the calculated values using the coupled-mode theory. (**D**) The SE decay curves at detunings of -1.0, 0, and 1.0 nm respectively, corresponding to the white curves in graph (A).

In a second set of experiments, the dynamic control of SE is achieved by replacing the thermo-optic tuning with free-carrier injection. In this case, a pulsed laser injects free carriers into the FP cavity. When the initial detuning between the target and FP-cavity is adjusted to be 0 nm, two coupled modes are observed at 1552.0 and 1552.4 nm. The laser pulse produces a blue-shift of the FP mode, bringing it out of resonance from the target cavity and the SE rate from QDs in the target cavity is enhanced due to the increase in the Q-factor and the increased localization of the vacuum field (Fig. 3A-C). This gives rise to a peak at time zero in the 3D map, which lasts until the FP mode relaxes to the initial wavelength due to diffusion and recombination of the free carriers. Note that the duration of this burst of SE (232 ps at full-width half-maximum (FWHM)) is related to the free-carrier lifetime in the FP cavity, and not to the emitter's lifetime – this shows the possibility of modulating SE at frequencies of several GHz, well above the bandwidth limitation related to the lifetime. This dynamic process is simulated by a master equation model, which reproduces the PL temporal dependence very well (Fig. 3C). The observed modulation depth at the peak, $\frac{I_{max}}{I_0} = 3.3$ results from the combined effect of LDOS enhancement, increase of photon population due to the Q change and change in collection efficiency due to the redistribution of the vacuum field.

For an opposite initial situation of nonzero detuning, the FP and target cavity modes can be transiently brought into resonance (Fig. 3D-F). This produces a sharp dip (246 ps at FWHM) in the SE intensity because of the LDOS reduction due to the vacuum field delocalization and Q-factor decrease, confirming that the peak in Fig. 3A is not due to the additional PL produced by the injected carriers. The measured peak modulation depth is $\frac{I_0}{I_{min}} = 2.0$ for this case. Note that the modulation occurs here over the entire mode spectrum, and clearly differs from the static modulation of the SE rate using PhC cavities [22], and the dynamic modulation which may be obtained by changing the emitter-cavity detuning by the control of cavity wavelength [23-24], or by Stark tuning of the exciton energy [9]. Indeed, our method directly changes the SE rate by controlling the interaction term $g$ and loss rate $\kappa$, which allows minimizing the chirp of the emitted photons and enables the ultrafast control beyond GHz, detrimental to application in quantum information processing[25].

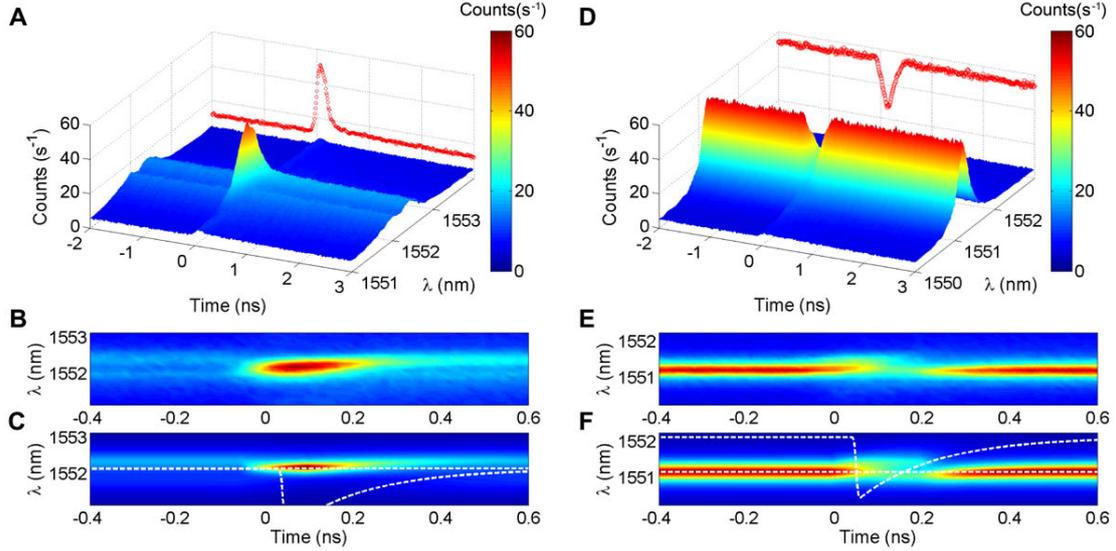

**Fig. 3.** Dynamic control of SE at the target cavity. (**A**) A time-resolved μPL map, for an initial detuning of 0 nm, made by measuring the decay curves at different wavelengths. The curve at the backplane is the measured μPL trace at the wavelength of 1552.2 nm. (**B**) The top view of the graph (A). (**C**) The simulation results compared to graph (B). The wavelengths of the target and FP modes are plotted in white dotted curves. (**D-F**) Same as (A-C) for an initial detuning of 0.6 nm. The curve at the backplane is the measured μPL trace at the wavelength of 1551.2 nm.

To further show the flexibility of our approach, we demonstrate the control of the temporal SE decay profile of the emitters in the target cavity. In this case both the target and the FP cavity are pumped by the same pulsed laser, with variable delay. The excitation power at the target cavity produces an initial blue shift in the target cavity mode. In order to avoid the effect of this blue-shift we choose delays such that the target cavity wavelength is stabilized. When the initial detuning is zero and the pulse exciting at the FP cavity is delayed by 2.0 ns, a spike is observed in the target cavity intensity when the FP mode is brought out of resonance (the inset of Fig. 4A). The appearing time of the spike can be control with various delay times of 1.5, 2.0 and 2.5 ns respectively (Fig. 4A). For the opposite situation when the static detuning between coupled modes is set to be 0.6 nm, the FP mode is transiently brought into resonance so that the expected dip appears (Fig. 4B).

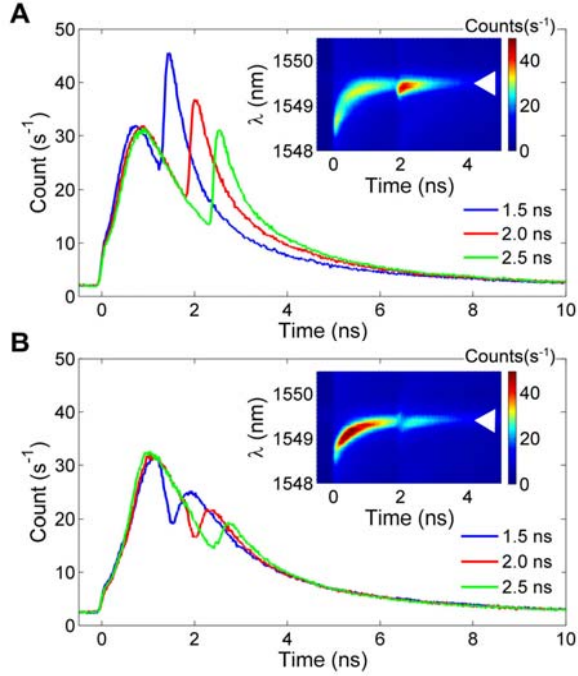

**Fig. 4.** Dynamic modulation of the SE decay profile at the target cavity. (**A**) μPL decay curves at the wavelength indicated by a white arrow in the inset, with different delay times of 1.0, 1.5, and 2.0 ns of the control pulse. The inset shows a time-resolved μPL map at zero initial detuning. (**B**) Same as (A) for an initial detuning of -0.6 nm.

In our demonstration, the temporal resolution is determined by the free carrier lifetime, which can be potentially reduced down to a few ps by applying an electric field in the control cavity[26]. This would allow shaping the control cavity frequency and thereby the vacuum field on ps timescales by the simple control of the pump pulse temporal profile. In the context of CQED, this ultrafast nonlocal control can be used to engineer the shape of single-photon pulses and for the on-off switching of Rabi oscillations, without directly disturbing the carrier population in the target cavity. As the mode field also determines the stimulated emission rate, the same method may be used to dynamically change the Q-factor and/or the gain in nanocavity lasers, leading to a new class of ultrafast Q- and gain-switching techniques for semiconductor lasers.

**Methods**

**Coupled mode theory.** The system of three coupled oscillators (the emitter, the target and FP cavities) can be described by a non-Hermitian Hamiltonian[27]

$$H = \hbar \begin{bmatrix} \omega_0 & g & 0 \\ g & \omega_t - i\kappa_t & \eta \\ 0 & \eta & \omega_{FP} - i\kappa_{FP} \end{bmatrix}, \quad (1)$$

where $\omega_0$, $\omega_t$ and $\omega_{FP}$ are the angular frequencies of the emitter, the uncoupled target and FP cavity modes respectively, $\kappa_t$ and $\kappa_{FP}$ are the corresponding loss rates. We assume that the emitter couples only to the target cavity with an interaction strength $g$, where we fix the interaction to the weak coupling regime $g \ll \kappa_t, \kappa_{FP}$. The coupling rate between the two cavities is $\eta$. By diagonalizing the 2x2 sub-matrix of the cavities, the normalized mode functions $E_{1,2}(r)$ of the coupled modes are calculated in the basis of the isolated cavity modes $E_t(r)$ and $E_{FP}(r)$, $E_{1,2}(r) = \alpha_{1,2} E_t(r) + \beta_{1,2} E_{FP}(r)$, with $\alpha_1 = \beta_2 = \alpha$, and $\alpha_2 = -\beta_1 = \beta$. In the basis of the coupled modes $E_{1,2}(r)$, the Hamiltonian is given by

$$H' = \hbar \begin{bmatrix} \omega_0 & \alpha g & \beta g \\ \alpha g & \tilde{\omega}_1 & 0 \\ \beta g & 0 & \tilde{\omega}_2 \end{bmatrix}, \quad (2)$$

where $\tilde{\omega}_{1,2} = \omega_{1,2} - i\kappa_{1,2}$ are the complex eigenvalues of the coupled modes. The effective dipole-coupled mode interaction rates $\alpha g$ and $\beta g$ and the loss rates $\kappa_{1,2}$ are functions of the detuning between the two cavities and can be controlled by tuning one of them. The SE rate into one of the coupled modes, normalized by the SE rate in the unperturbed target cavity is given by $\frac{\gamma_1}{\gamma_t} \cong |\alpha|^2 \frac{\kappa_t}{\kappa_1}$, showing that SE is affected both by the change in the distribution of the vacuum field (or equivalently the mode volume) (term $|\alpha|^2$) and by the modulation of the Q-factor. The thermo-optic tuning results in Fig. 2 have been fitted by diagonalising the Hamiltonian (1), while the free-carrier tuning data of Fig. 3 has been modeled solving the master equation corresponding to Eq. (1), with the addition of incoherent pumping.

**Sample preparation.** The sample was grown on an InP (100) substrate by metal-organic vapor phase epitaxy. The structure contains a 100 nm-thick InP buffer layer, followed by a 110 nm lattice matched InGaAsP layer with a bandgap at 0.992 eV (Q1.25), 1.2-monolayer-thick GaAs interlayer, a single layer of InAs QDs, a 110 nm InGaAsP layer, and a 50 nm InP capping layer. The QDs have an areal density of $2 \times 10^9$ cm$^{-2}$ and provide a 100 nm broad luminescence peak around 1550 nm[28], which feeds the cavity mode. The PhC cavities were fabricated with a standard process by electron beam lithography and inductively coupled plasma using Cl$_2$/Ar/H$_2$ mixture. The selective wet etching of the InP sacrificial layer was done in a HCl/H$_2$O solution at 2°C. The lattice constant of PhC is chosen to be $a_0 = 480$ nm with a filling factor of 0.30 to achieve a cavity mode around 1550 nm at 77K. The target and FP cavity consist of 2 and 80 periods of modulated lattice constant ($a_1 = 1.03 \times a_0$), respectively. The barrier between two cavities contains four periods of the original lattice constant.

**Experimental set-up.** The measurement is performed at 77K using a confocal microscopy setup, where two laser beams are focused at different positions with a separation of 30 μm, as indicated in the sketch in Fig. 1. The micro-photoluminescence (μPL) signal from the target cavity is collected by the objective and measured with a spectrometer. The experiments on thermo-optic tuning (Fig. 2) were performed by exciting the target cavity with a pulsed laser at 1064 nm with a pulse width of 6 ps and energy of 2 μJ/cm$^2$, while the FP cavity was heated with a continuous wave (CW) beam at 780 nm. Decay curves at different detunings were measured by time-correlated single-photon counting (TCSPC) using a superconducting single-photon detector by filtering out the cavity peak with a narrow bandpass filter (FWHM=0.5 nm)[29]. In the ultrafast tuning experiments (Fig. 3) the target cavity was pumped with the CW beam at 780 nm while the FP cavity was excited with the 1064 nm pulsed laser with energy of 200 $\mu$J/cm$^2$, and the time-resolved PL from the target cavity was measured by TCSPC. In the dynamic modulation experiment of Fig. 4 both the target and the FP cavities were excited with the pulsed laser, at different delays, with energies of 20 and 200 $\mu$J/cm$^2$, respectively. The fine adjustment of the wavelength detuning between two cavities was achieved by coupling a CW lasing beam at 780 nm on the FP cavity.

**Acknowledgments**

The authors are grateful to B. Wang, M.A. Dündar, and R.W. van der Heijden for fruitful discussions, to Z. Zhou, D. Sahin, F.M. Pagliano, C.P. Dietrich, E.J. Geluk, E. Smalbrugge, T. de Vries, M. van Vlokhoven, J.M. van Ruijven, and P.A.M. Nouwens for technical support, and to P.M. Koenraad and E. Pelucchi for a critical reading of the manuscript. This research is financially supported by NanoNextNL, a micro and nanotechnology program of the Dutch ministry of economic affairs, agriculture and innovation (EL&I) and 130 partners, the Dutch Technology Foundation STW, applied science division of NWO, the Technology Program of the Ministry of Economic Affairs under project No. 10380 and the FOM project No. 09PR2675.